\documentstyle[aps,pra,psfig,multicol]{revtex}

\begin{document}
\title{Electrical current distribution across a metal-insulator-metal structure
during bistable switching }
\author{C. Rossel,\thanks{To whom correspondence should be addressed;
e-mail: rsl@zurich.ibm.com} G. I. Meijer, D. Br\'emaud, and D. Widmer}
\address{IBM Research, Zurich Research Laboratory, CH-8803 R\"{u}schlikon,
Switzerland}

\draft
\maketitle

\begin{abstract}
Combining scanning electron microscopy (SEM) and
electron-beam-induced current (EBIC) imaging with transport
measurements, it is shown that the current flowing across a
two-terminal oxide-based capacitor-like structure is
preferentially confined in areas localized at defects. As the
thin-film device switches between two different resistance
states, the distribution and intensity of the current paths,
appearing as bright spots, change. This implies that switching
and memory effects are mainly determined by the conducting
properties along such paths. A model based on the storage and
release of charge carriers within the insulator seems adequate to
explain the observed memory effect.
\end{abstract}

\begin{multicols}{2}

\section{Introduction}
\vspace*{-4mm}

The use of thin-film heterostructures for making two-terminal
switching and memory devices has been a long-standing challenge.
Much work has been done in the field, starting already in the
late sixties on various materials such as amorphous
semiconductors \cite{Ovshinsky,Neale,Dewald}, ZnSe-Ge
heterojunctions \cite{Hovel} as well as on several binary oxides
\cite{Hiatt,Hickmott,Chopra,Argall,Gibbons} to produce simple
crosspoint memory arrays with bit and word lines. More recently,
work on porous Si \cite{Ueno} and polymers \cite{Gao,Liu,Krieger}
has also demonstrated electrical bistability. The enhanced
interest in perovskites since the discovery of high-temperature
superconductivity in the cuprates and the giant and colossal
magnetoresistance in the manganites has generated a tremendous
effort in the investigation of oxide thin films. In particular
the replacement of SiO$_{2}$ as gate oxides in CMOS technology
and the fabrication of smaller ferroelectric memories (FRAM) have
triggered the research of new, alternative materials. Earlier, we
reported the observation of reproducible switching and memory in
Cr-doped (Ba,Sr)TiO$_{3}$ and SrZrO$_{3}$ epitaxial-film-based
metal-insulator-metal (MIM) devices \cite{Beck}. In particular it
was shown that not only was reversible switching between a
low-resistance and a high-resistance state possible but also that
multilevel switching could be achieved. In addition, a
retention-time test performed over 12 months demonstrated  the
nonvolatile memory properties of these memory cells. Nevertheless
an important issue is the scalability of such devices to lower
dimensions and the influence of intrinsic film defects on the
current flow and the switching mechanism itself. Here we report
on the simultaneous investigation of the film microstructure by
scanning electron microscopy (SEM), the distribution of the
electrical current flow by electron-beam-induced current (EBIC),
and the switching properties of individual memory cells.

\vspace*{-4mm}
\section{Experimental}

\vspace*{-4mm}
\subsection{Oxide thin films and heterostructures}

\begin{figure}
\begin{center}
\narrowtext
\leavevmode
\psfig{figure=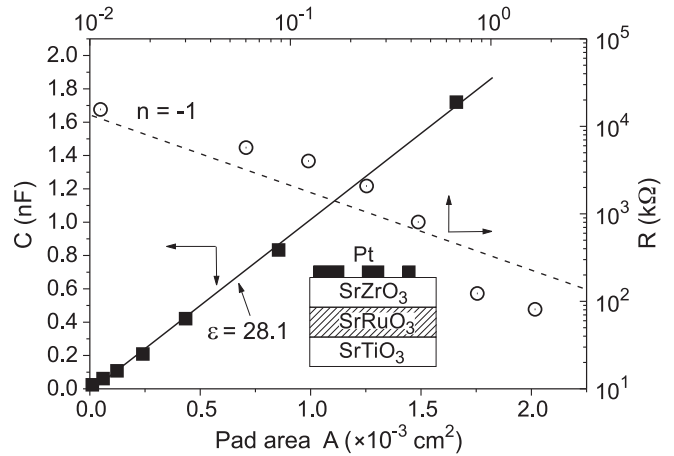,width=86mm}
\end{center}
\caption{Capacitance and resistance (log-log plot)
versus pad area of a Pt/SrZrO$_{3}$(0.2\%
Cr)/SrRuO$_{3}$/SrTiO$_{3}(001)$ MIM structure at 300 K. The
SrZrO$_{3}$ film is 35 nm thick. Both dependences point to a good
scaling down to the smallest pad size, 40 $\mu $m in diameter.}
\label{fig1}
\end{figure}

The MIM structure is shown in the inset of Fig.\ \ref{fig1}. As a
base metallic electrode a SrRuO$_{3}$ film of about 4 nm  is
grown epitaxially on SrTiO$_{3} $ (001) by pulsed laser
deposition (PLD) at 700$^{\circ }$C and an O$_{2}$ pressure of
250 mTorr using a KrF excimer laser. The insulating oxide film is
then deposited in situ at a slightly higher temperature in a
typical thickness of approx.\ 35 nm as determined  by X-ray
diffraction from the finite-size oscillations observed at low
angles and at the main Bragg reflections. We focus here on
SrZrO$_{3}$ films doped with 0.2 at.\% Cr (i.e., $3\times
10^{19}$ at/cm$^{3}$), which were found earlier to yield good
switching properties \cite{Beck}. The top electrodes of Pt are
evaporated by electron beam through a shadow mask to define
squares or disks of variable sizes, ranging between 460 and 35
$\mu$m, which corresponds to areas between about 2$\times
$10$^{-3}$ and 10$^{-5}$ cm$^{2}$. The epitaxy of the films was
checked by X-ray diffraction, the surface topography by atomic
force microscopy (AFM) imaging. Transport measurements were
performed either in ac mode with a LCR meter HP4263 or in dc mode
using a Keithley 238 current source measuring unit run in
voltage- or current-controlled mode. Voltage is applied to the
top electrode via a CuBe-coated tungsten needle. A current
limitation is set to prevent breakdown and destruction of the
structures. A typical dependence of the resistance and
capacitance as a function of pad area $A$ is shown in Fig.\
\ref{fig1} for a 35-nm-thick SrZrO$_{3}$(0.2\% Cr) film. In this
range of pad sizes, the resistance largely follows the $ R
\propto A^{-1}$ relation as shown on a log-log plot, whereas the
capacitance increases linearly with $A$, leading to a dielectric
constant of about 28.

From these scaling properties, it appears that the film transport
properties are rather homogeneous down to a length scale of about
50 $\mu $m. Below, they become more strongly dependent on the
distribution of microstructural defects such as grain boundaries,
precipitates of other phases or micro pinholes, and a systematic
scaling analysis is more difficult. Pad sizes down to 100 nm were
successfully patterned on thinner films by electron-beam
lithography and lift-off techniques \cite{Wind} and investigated
by AFM with a conducting tip \cite{Bietsch}. Switching by voltage
pulses was observed, but a systematic analysis was difficult
because of local inhomogeneities and varying sticking properties
of the metal pads.

\vspace*{-4mm}
\subsection{Electron-beam-induced current method}

\begin{figure}
\begin{center}
\leavevmode
\psfig{figure=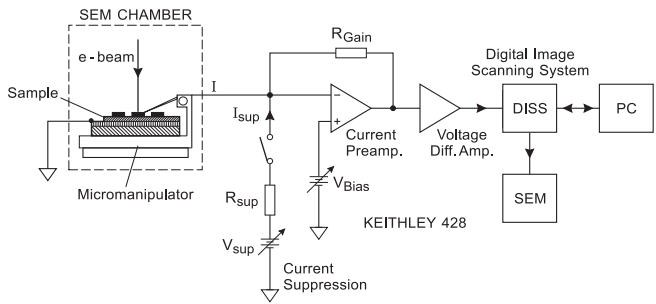,width=86mm}
\end{center}
\caption{Schematics of the EBIC measurement system.}
\label{fig2}
\end{figure}

EBIC has primarily been used to investigate the electrical
properties of metal-oxide-semiconductor (MOS) capacitors,
focussing on the leakage and breakdown properties of oxide thin
films grown on Si \cite{Lin,Kirk,Tamatsuka,Lau,Lin2}. In those
cases an internal electrical field allows the specimen to act as
detector of the charge carriers generated by the incident
electron beam. The injected carriers are subject to drift,
diffusion, trapping and recombination events. The formation of a
space charge region at the interface with the bulk Si ($p$-type
or $n$-type) separates the electron-hole pairs created by each
absorbed electron, resulting in a large current gain \cite{Kirk}.
The output signal can be amplified by several orders of
magnitude. Performing EBIC in a SEM provides a way to map and
quantify the inhomogeneities in the electrical properties of a
material that are due to several types of intrinsic defects. Since
the pioneering work of Pensak \cite{Pensak}, to our knowledge,
very few EBIC studies have been done on MIM structures, such as
ours, using an insulating oxide epitaxially grown on a metallic
oxide base electrode. The diagram of the experimental arrangement
is shown in Fig.\ \ref{fig2}. A JEOL 6400 scanning microscope has
been equipped with a custom-made removable micromanipulator. The
electrical contact to the top electrode is done with a
CuBe-coated tungsten needle. Bias polarity is defined with
reference to the SrRuO$_3$ bottom electrode connected with silver
paint to a thin Cu wire. The low-noise EBIC amplifier (Keithley
428) is connected to the sample in the vacuum chamber by coaxial
feedthroughs. A useful aspect of the circuit is the current
suppression option, which enables  subtraction of the background
signal. The data-acquisition system consists of a digital imaging
scanning unit (WinDiss-2 from Point Electronic GmbH, Halle,
Germany) with two analog 12-bit input channels. For the
measurements, the electron-beam energies $E_{b}$ were varied
between 20 and 30 kV with beam currents $I_{b}$ ranging from 0.1
to 7 nA, as verified by a Faraday cup.

\vspace*{-4mm}
\section{Results}

\begin{figure}
\begin{center}
\leavevmode
\psfig{figure=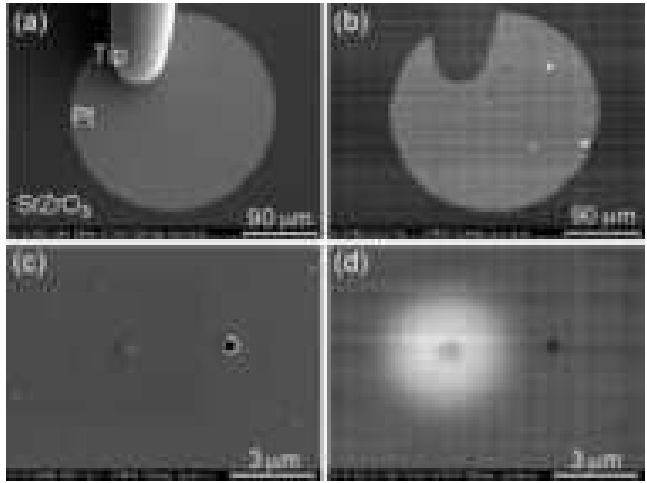,width=85mm}
\end{center}
\caption{(a) SEM image of a Pt/SrZrO$_{3}$(0.2\% Cr)/SrRuO$_{3}$
MIM capacitor structure with Pt electrode, 250 $\mu $m in diameter
and 100 nm thick. (b) Corresponding EBIC image at $V_{b}=0\ $V with
the appearance of conducting white spots. Beam-acceleration
voltage $V_{\rm acc}=25$ kV. (c) Enlarged SEM view showing a defect in
the film below the Pt electrode and a hole in this electrode, and (d)
corresponding EBIC image with one bright leakage spot correlated
to the defect but not to the hole.}
\label{fig3}
\end{figure}

The first EBIC measurements coupled to electrical switching were
done using rather thick, 100-nm Pt top electrodes to ensure that
the tip did not scratch or perforate the SrZrO$_{3}$ oxide layer
(same film as for Fig.\ \ref{fig1}). In all  memory cells
investigated, the EBIC images taken in the high-resistive state
systematically showed a given number of white spots corresponding
to discrete paths of flowing current (Fig.\ \ref{fig3}). Most of
the pictures are taken at small ($< 1$ V) or zero voltage because
the application of larger bias up to $\pm 5$ V lead to enhanced
background noise without changing the image features. As
expected, a change of the bias polarity produced an inversion of
the image contrast. Even when displacing the needle slightly, no
change in the flow configuration was observed and no visible
current path was generated by the pressure of the needle at the
contact location. Thus the main current flow seems to be
restricted around some of the defects in the SrZrO$_{3}$ film,
but not all defects contribute to it. This point is supported by
Fig.\ \ref{fig3}(c,d), which is an enlarged view of the Pt
electrode surface. Next to a small circular hole in the Pt, 350
nm in diameter, a protuberance of similar size pops up because of
an underlying defect in the oxide film [Fig.\ \ref{fig3}(c)]. The
corresponding EBIC image [Fig.\ \ref{fig3}(d)] shows that the
current flows only on that defect and not through the hole. Other
smaller defects and visible surface particles, possibly dust, do
not generate any electron-beam-induced currents at this
sensitivity scale. Because of diffused electrons in the various
layers, the radius of the white spot is  not a uniquely defined
quantity here as it depends on the amplification gain and on
intensity and contrast factors of the scanned image. At low
current densities, a lateral resolution of 20 nm can be achieved
with our system.

An interesting observation was made when focusing the entire
electron beam on one of the conducting spots. By applying the
maximum beam current ($I_{b}= 18$ nA) at 35 kV for successive
periods of time (1 to 30 min), a series of SEM/EBIC images were
taken under the same conditions ($E_{b} = 25$ kV; $I_{b} = 1.5$
nA). Surprisingly, the intensity (and radius) of the spot
decreased progressively until it became almost invisible, whereas
the corresponding resistance of the MIM structure measured at 0.2
V increased by about 8\% from 730 to approx.\ 790 k$\Omega$. The
direct influence of a carbon film grown by the incident beam
should be negligible as long as the underlying Pt layer is not
removed or modified. We found in addition that not all the spots
treated in that manner could be made to fully disappear. A
possible reason for this extinction effect will be discussed
below.

\vspace*{-4mm}
\subsection{Bistable switching and EBIC}

\begin{figure}
\begin{center}
\leavevmode
\psfig{figure=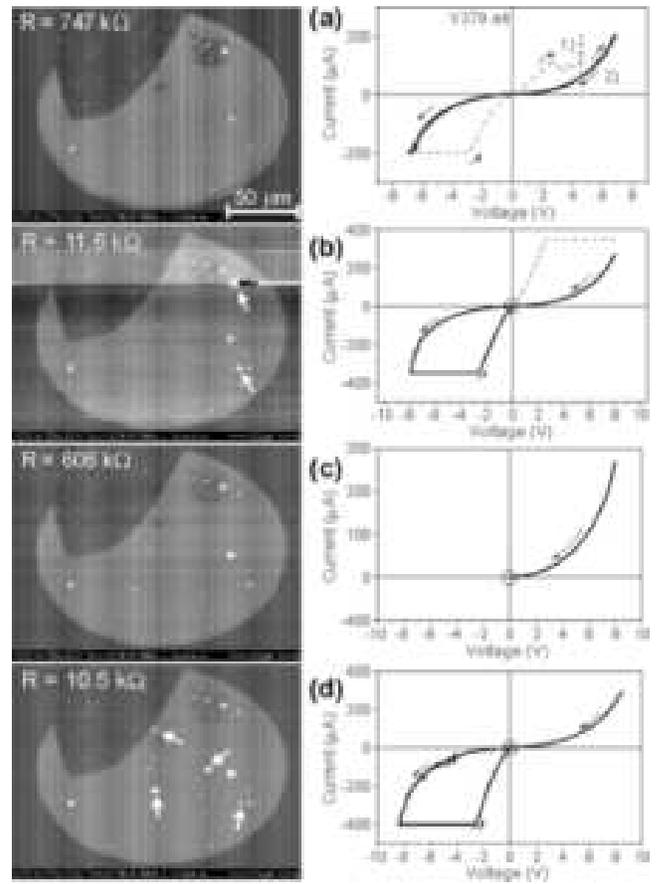,width=85mm}
\end{center}
\caption{Sequence of EBIC images and the corresponding $I$--$V$
characteristics for a Pt/SrZrO$_{3}$(0.2\% Cr)/SrRuO$_{3}$ memory
cell, 175 $\mu$m in diameter. Pt electrode thickness is 5 nm.
$V_{b}=0$\ V, $V_{\rm acc}=25$ kV. (a) Initial high-impedance
state. $I$--$V$ curve 1:\ partial switching has occurred, curve 2
the sample has not yet switched. (b) After one cycle down to --8
V and up to 0 V, the cell has switched to its low-impedance
state, and two new spots appear in the EBIC image (arrows). (c)
The cycle is completed up to 8 V, the cell switches back to its
high-impedance state at 0 V, and the two spots disappear. (d)
Return to the high-impedance state after full cycle sweep. New
spots appear (arrows).} \label{fig4}
\end{figure}

Let us now turn to the correspondence between electrical switching
and the EBIC images. For this purpose we have grown similar films
with thinner (4--5 nm ) top Pt electrodes. Under these
conditions, the microstructure and defects of the films under the
metal electrode can be seen with surprisingly good contrast and
accuracy. A sequence of four EBIC pictures is displayed in Fig.\
\ref{fig4} with the corresponding $I$--$V$ characteristics
defining the resistance state of a memory cell 175 $\mu$m in
diameter. In the initial high-resistance state, typically 750
k$\Omega$, one observes about 10 white spots of various sizes
distributed across the electrode in the EBIC picture [Fig.\
\ref{fig4}(a)]. In this stage, a set of symmetric and reversible
$I$--$V$ characteristics was taken at a rate of 0.1 to 0.2 V/s,
starting from the maximum positive bias voltage, $V_{\max}$.  By
progressively increasing $V_{\max}$, the cell started to switch
partially for a maximum bias of 7 V and a compliance current of
200 $\mu$A, as seen from the dashed curve 1 in Fig.\
\ref{fig4}(a). From the respective slopes at zero bias, a
resistance ratio $r = R_{\rm high}/R_{\rm low}$ = 22 was derived.
In the following run, the sample, which was not completely
preformed, returned to its high-resistance state (reversible
curve 2). When increasing the voltage to 8 V and the limiting
compliance to 350 $\mu$A, bistable and reproducible switching set
in. The next run was then stopped at 0 V after sweeping to
negative voltage and switching the memory cell into its
low-resistance state ($R_{\rm low}=11.6$ k$\Omega$) [Fig.\
\ref{fig4}(b)]. In that state a new EBIC image was scanned
showing the clear appearance of two new intense spots (see
arrows). The intensity of the other spots remained practically
unchanged. As the device was switched back to its high-resistance
state (606 k$\Omega$) after a sweep up to 8 V and down to 0 V,
the two spots disappeared almost completely [Fig.\
\ref{fig4}(c)]. At the same time some of the existing spots became
slightly more intense. To check whether the current-path
distribution was reversible, the cell was switched back to its
low-resistance state ($R_{\rm low}=10.5$ k$\Omega$) through an
entire cycle so as to start the sweep again from positive bias
voltage. Surprisingly, only one of the two previous spots
reappeared, accompanied by three new intense spots [arrows, Fig.\
\ref{fig4}(d)]. This observation leads us to conclude that
electrical switching in thin films correlates to a large extent
to the properties of individual conducting paths that can be
locally turned on and off or changed in intensity. These paths
are located at the positions of intrinsic defects in the film.
Nevertheless at low current amplitudes or in a very high
resistive state, an homogeneous contribution of the bulk of the
film cannot be ruled out. In the present case, if one assumes
that the entire current of 400 $\mu$A flows along about 15
channels having an average area of 10 $\mu$m$^2$, we get a rather
high current density of $4 \times 10^3$ A/cm$^{2}$.

\begin{figure}
\begin{center}
\leavevmode
\psfig{figure=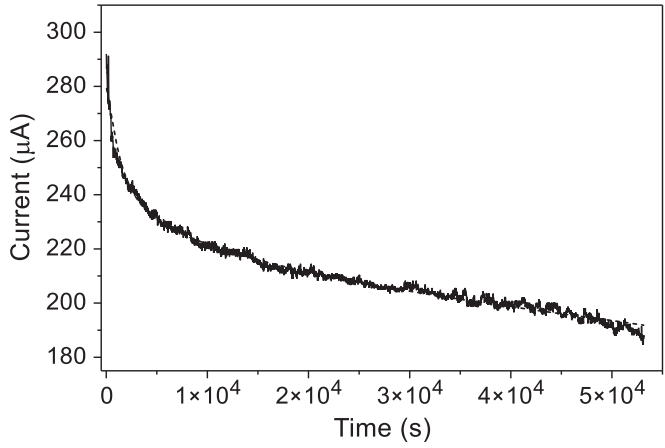,width=87mm}
\end{center}
\caption{ Current relaxation $I(t)$ as the memory cell is
progressively returned to its higher impedance state under a
stress voltage of 9 V. The data are well fitted with a double
exponential decay law (dotted line).} \label{fig5}
\end{figure}

After this treatment the cell was switched back and forth many
times, keeping it either in the low- or the high-resistance state
at zero bias voltage for up to 48 h without any change,
confirming the nonvolatility of the given state. In the dc mode,
the switching ratio $r = R_{\rm high}/R_{\rm low}$ was found to
vary between approx.\ 45 and 90, depending on the amplitude of
the compliance current set to prevent breakdown. Fast switching
was also performed by application of short pulses for writing and
erasing, as described in \cite{Beck}. In this mode we observed
that for a given voltage pulse amplitude, the ratio $r$ increases
continuously with the pulse length between 1 ms and 1 s, i.e.,
the total number of charges injected into the insulator plays the
dominant role in the switching mechanism. After several
experiments at increasingly higher current values, the memory
cell become locked into its low-resistance state. Nevertheless, by
applying a constant voltage of 9 V for a longer time, the cell
can be repaired and partially restored into its high resistive
state. As shown in Fig.\ \ref{fig5}, the current measured over a
period of 14 h decreased continuously, with a change in
resistance from 25 to 118 k$\Omega$ measured at 0.2 V. This time
dependence can be well fitted by a double exponential decay
function of the form $I(t)=I_{0}+A_{1}\exp (-t/\tau
_{1})+A_{2}\exp (-t/\tau _{2})$, with long time constants $\tau
_{1}= 0.52$ h and $\tau _{2}= 11.8$ h. At that stage the device
was stable and could  again be switched with pulses. This means
that the conducting paths have been modified by this forming
process.

\vspace*{-4mm}
\subsection{Microstructural defects}

The important question that arises at this point is the nature of
the defects and their individual electronic properties. If
bistable switching is by nature related to the conduction
processes of one or several local paths, it would be favorable to
control their number or artificially recreate their structure.
Such work is under way on specially dimpled single crystals of
Cr-doped SrTiO$_{3}$.

\begin{figure}
\begin{center}
\leavevmode
\psfig{figure=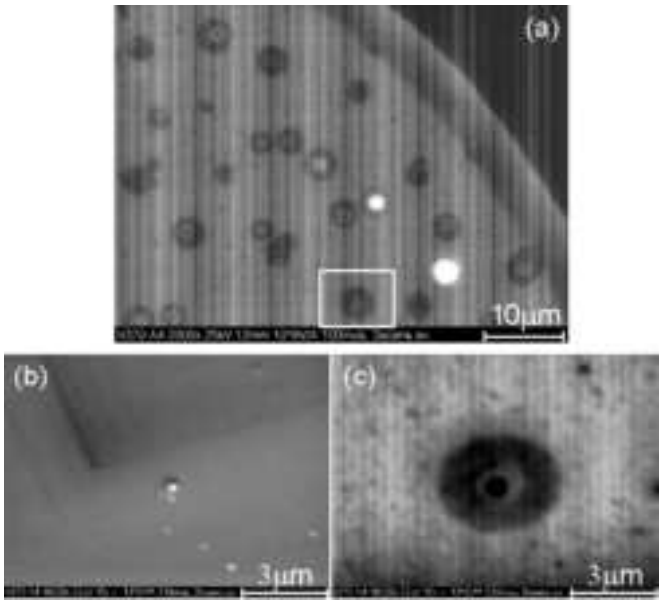,width=87mm}
\end{center}
\caption{ (a) Enlarged EBIC image of the switching device
investigated in Fig.\ \protect\ref{fig4}, displaying two bright
leakage spots and other dark spots of similar geometry
corresponding to defects localized in the SrZrO$_{3}$ film;
$V_{b}=0$ V, $V_{\rm acc}=25$ kV. (b) SEM  and (c) close-up EBIC
images of the rectangular area taken under a tilt angle of
37$^{\circ}$, showing the main defect and its transmitted current
mapping. Sharp contrasted microstructural details of the film
under the Pt layer become apparent.} \label{fig6}
\end{figure}

\begin{figure}
\begin{center}
\leavevmode
\psfig{figure=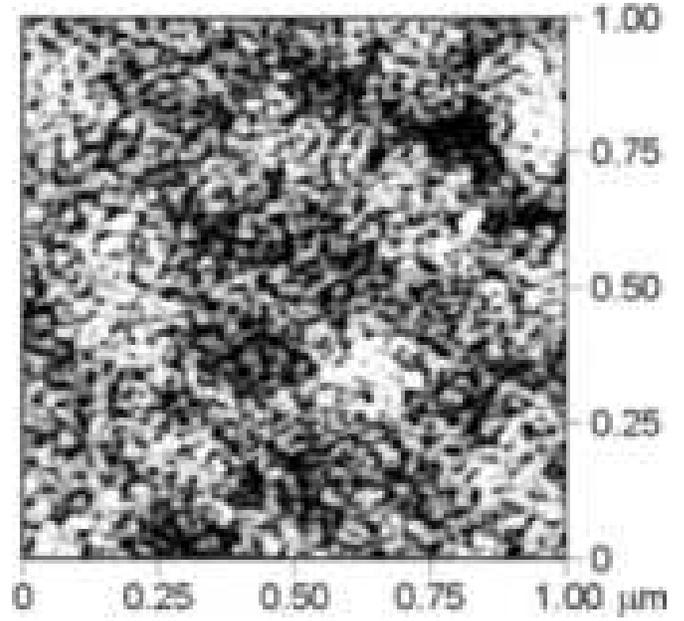,width=87mm}
\end{center}
\caption{
AFM image of the SrZrO$_{3}$(0.2\% Cr) film grown on SrRuO$_{3}$/SrTiO$_{3}(001)$
investigated in Figs.\ \protect\ref{fig4} to \protect\ref{fig6}. Greyscale
amplitude = 6.2 nm.}
\label{fig7}
\end{figure}

EBIC images can strongly enhance microstructural details that are
not visible in conventional SEM images of collected secondary
electrons. This contrast enhancement is true for MOS devices
\cite{Kirk}, but especially so in MIM structures. As an example
two enlarged sections of the surface of the memory cell measured
above are displayed in Fig.\ \ref{fig6}. The EBIC image of Fig.\
\ref{fig6}(a) shows the Pt electrode (5-nm thick), its rounded
edge with decreasing thickness (shadow effect) and the
SrZrO$_{3}$ film underneath. Next to the two white conducting
spots, a number of black dots, smaller than 500 nm in diameter and
surrounded by a dark halo (2--5 $\mu$m), are scattered across the
Pt surface. These features are only seen for a thin Pt layer and
correspond to microstructural defects in the oxide film; some of
them, but not all, are observable in the SEM image. Two zoomed-in
views corresponding to the white rectangle and taken under a tilt
angle of 37$^{\circ }$ are displayed in Fig.\ \ref{fig6}(b) and
(c). The SEM picture shows a smooth surface with the presence of a
main dot (250 nm in diameter) at the center and a few smaller
ones scattered around it. In the corresponding EBIC image, the
main defect is clearly visible with its surrounding cloud, but,
more interestingly, the granularity of the nanostructure of the
film below the platinum electrode becomes apparent with features
having a typical size of 200 nm or less, which corresponds to the
topography found by AFM (Fig.\ \ref{fig7}). Indeed, next to the
fine nanostructure displayed in the AFM image, bright and darker
regions defining hillocks and dips of about 200 nm are in evidence
over the scanned area ($1 \times 1$ $\mu$m$^{2}$). The origin of
the defects seen in Fig.\ \ref{fig6} can be partly attributed to
the LPD technique itself \cite{Willmott}. Besides the many
advantages of laser ablation, some of its inherent drawbacks are
the projection of microscopic ejecta during the ablation process,
which can stick to the substrate, the presence of microditches or
holes caused by bombardment with high-kinetic-energy particles,
and the insertion of impurities present in the target material.
In all these cases it cannot be excluded that their presence in
the film locally creates a microstructural stress field that only
becomes apparent with methods based on electron- or light-beam
injection. Another explanation for the clouds in the EBIC scans
could be a local modification of the interface properties between
the metal electrode and the film, such as a local change in the
respective work functions due to oxidation of the metal electrode
and depletion of the oxide layer.

In thicker films of the same manufacture, the number of
observable white spots, such as those seen in Figs.\ \ref{fig3} or
\ref{fig4}, is less important at small bias voltage but definitely
increases at higher voltages. Once the compliance is released and
the current progressively raised until thermal breakdown occurs,
localized ``burned'' pits with crater-like geometry are generated
and the structure is shorted. The interesting observation is that
their number is larger than the EBIC spots observed in the
prebreakdown regime but all EBIC spots correspond to a ``burned''
channel. The general coincidence of breakdown sites with defects
such as stacking faults and the contrast mechanisms at these
sites have been studied by many researchers but almost
exclusively in gate oxides grown on Si
\cite{Lin,Kirk,Tamatsuka,Lau,Lin2}.

\vspace*{-4mm}
\section{Discussion}

A well-known observation in dielectric ceramics is the resistance
degradation induced by dc electrical field stress
\cite{Schaumburg,Odwyer}. Such a degradation, characterized by a
slow increase of the leakage currents, may occur at temperatures
and fields much below the onset values of dielectric or thermal
breakdown. This wear-out process, which is the main limiting
factor for the lifetime of multilayer ceramic capacitors, is not
destructive and reversible upon annealing. From the large number
of studies on oxides and in particular on perovskite ceramics
such as SrTiO$_{3}$ \cite{Schaumburg,Waser1,Waser2,Dietz}, it
appears that the degradation rate is largely influenced by grain
size, stoichiometry, doping, second phases, porosity, and
electrode properties. In general, doping with donors such as
La$^{3+}$, which goes into the A sites (e.g., Sr$^{2+}$) of the
ABO$_{3}$ perovskite lattice, leads to an increase of the leakage
current and improves or stabilizes the degradation \cite{Hofman}.
On the other hand, acceptor dopants such as Al$^{3+}$, Fe$^{3+}$,
Mn$^{3+}$, or Cr$^{3+}$ \cite{Hofman,Paek}, which partially
substitute on the B site (e.g.\ Ti$^{4+}$), lead to a significant
decrease in the leakage and relaxation currents and to a
suppression of the degradation rate. As mentioned by Waser {\it
et al.} \cite{Waser1,Waser2} the current response of high-$K$
thin films under a voltage step excitation exhibits three
distinct regimes: a short-term relaxation regime with a power-law
decay $J \propto t^{-\alpha }$ ($\alpha \simeq 1$) attributed to
bulk microstructural effects (i.e., degree of crystallinity), a
mid-term leakage regime with little change in $J$ (electrode
interface and bulk-doping effects), followed by a resistance
degradation regime with increasing $J$. This last regime, which
is well understood in titanate single crystals \cite{Baiatu}, is
more complicated in thin films and seems to be induced both by
electronic and ionic transport. Clearly, the double exponential
decay law that we observe in the recovery phase of our film in
Fig.\ \ref{fig5} under rather high electrical fields ($E=2.6
\times 10^{8}$ V/m) is different from the above-mentioned power
law valid for a pristine film at lower field. This behavior would
point towards a different mechanism dominated by the flow of the
current in localized channels rather than homogeneously across
the bulk of the film.

If this idea is correct, as our EBIC experiment indicates, it is
to be expected that the mechanism for switching and memory is also
dominated by the properties of such channels or filaments. One
reason is that switching never occurs without forming the
material, i.e., reducing its resistance from the
M$\Omega$--G$\Omega$ range to the k$\Omega$--M$\Omega$ range.
During forming, the weaker regions of the dielectric undergo an
enhanced electric-field stress and become more suitable for the
flow of current. This current is mainly due to the hopping or
tunneling of charges between adjacent sites or traps localized in
the gap of the insulator. A possible mechanism can be related to
the filling and defilling of traps by impact or field ionization
as proposed by several authors \cite{Hovel}. One very appealing
model is that of Simmons and Verderber \cite{Simmons} proposed
earlier to explain voltage-controlled negative resistance and
reversible memory effects in thin films of SiO sandwiched between
Al and Au electrodes. The idea is that a broad band of localized
states is created in the insulating gap by doping and forming. In
our case we can expect that the doping with Cr$^{3+/4+}$ ions ($3
\times 10^{19}$/cm$^{3}$) creates such sites separated by about 3
nm, in addition to the intrinsic traps. This separation is short
enough to favor a tunneling process for the transport of electrons
injected from the cathode. Assuming that the occupation
probability of these sites is determined by the kinetics of
tunneling rather than thermal statistics (small Boltzmann term)
and that the energy of the electron is conserved during tunnel
transition, Simmons and
Verderber \cite{Simmons} suggested an $I$--$V$
curve of the form $I=a \, \sinh (bV)$ for the low-impedance
state. The coefficient $a=A\exp (-1.02s(\phi
_{0}-\frac{1}{2}E_{0}s)^{1/2}$ is a transition probability, where
$A$ is a constant, $\phi _{0}$ the metal-insulator barrier
height, $s$ the average intersite spacing, and
$E_{0}=-Q_{0}/K\varepsilon _{0}$ the electrical field at the
cathode-film interface derived from Poisson's equation. $Q_0$ is
the total uncompensated charge in the depletion region of the
insulator.
The second factor is $b=0.13s^{2}/(\phi _{0}-\frac{1}{2}%
E_{0}s)^{1/2}.$ To explain the memory effect, characterized by
the presence of an hysteresis loop in the $I$--$V$ cycle, it is
assumed that charges can be stored in the insulator when the
applied voltage exceeds a certain threshold equivalent to the
energy $\phi _{i}$ defining the top of the localized levels with
respect to the Fermi level. The storage of negative charges
$qN_{s}$ near the top of the localized band in the insulator
reduces the electrical field at the interface to $E_{0}^{\prime
}=-(Q_{0}-qN_{s}/2)/K\varepsilon _{0}$, and consequently the
current passing through the system is lower, according to the
above equation for $I(V)$, in which $E_{0}$ is replaced by
$E_{0}^{\prime }$. Thus a higher impedance state is reached by
storing trapped charges, whereas the lower impedance state is
reestablished by releasing them. The ratio of the two impedances
at $V=0$ is therefore given by the amount of charge that can be
stored in the system. This also explains why multilevel switching
can be achieved in our thin films by the application of voltage
pulses of different heights \cite{Beck}. The relevant quantity is
in fact the amplitude of the current pulse times its width, which
determines the total amount of charge storage. According to our
observations, it appears that if the pulses are too short for a
given amplitude or if the interval between two successive pulses
is too small, the memory effect tends to vanish. A plausible
explanation for the observed dead time is that the stored charges
need a finite time to migrate to the center of the insulator,
possibly by a phonon-assisted tunneling process, before they can
escape.

To check the validity of this model at room temperature, we have
fitted the $I$--$V$ curves corresponding to the high- and
low-impedance states of Fig.\ \ref{fig4} with the above equation
$I=a\, \sinh (bV)$. It is found that the $I$--$V$ curves of both
states can be very well adjusted to the $\sinh(V)$ dependences,
in support of the model. We have also compared our fits with
other expressions describing carrier transport in insulators,
such as those for Schottky, Frenkel--Poole and Fowler--Nordheim
field emissions, as well as space-charge-limited conduction
\cite{Odwyer,Sze}. It appears that the high-impedance $I$--$V$
curves can also be fitted with the equation $I=\alpha V\,
\exp(\beta \sqrt{V}/T-q\phi _{0}/kT)$ known for Frenkel--Poole
emission, where the conduction is due to field-enhanced thermal
excitation of trapped electrons into the conduction band. Here
$\alpha $ and $\beta $ are two constants. Nevertheless additional
attempts with the expression derived for the tunnel or field
emission in the Fowler--Nordheim regime, i.e., $I=\alpha V^{2}\,
\exp(-\beta /V)$, failed. In that case the tunnel emission is
caused by field ionization of trapped electrons into the
conduction band or by tunneling from the metal Fermi level into
the insulator conduction band. This mechanism is expected to apply
only at higher voltages, above the threshold at which the memory
cell switches. Concerning the low-impedance state, satisfactory
fits could only be achieved with the Frenkel--Poole model or with
an expression of the type $I=\alpha V+\beta V^{2}$. Here the
ohmic dependence at low field is the result of thermally excited
charges hopping between adjacent isolated states, whereas the
quadratic term corresponds to a space-charge-limited current
resulting from the injection of carriers into an insulator with
no compensating charges. Based on the analysis of the $I$--$V$
curves in both impedance states, it seems that several transport
mechanisms based on hopping or tunneling through the insulator
are possible, but the model of Simmons and Verderber is very
appealing because it agrees with our data and provides a good
explanation of the nonvolatile memory effect in terms of the
storage and release of charges. It even provides an explanation
of multilevel switching and the various requirements for the
pulse amplitudes, widths and sequences. Further investigations in
particular on the temperature dependence of the switching and
memory effect, are needed to confirm or refine the model. Indeed,
we know that the switching effect is temperature dependent and
tends to vanish around 400 K. Current measured in both high- and
low-impedance states down to 77 K exhibits a thermally activated
behavior \cite{Beck}. In view of our EBIC results on films, where
the filamentary distribution of the electrical current has been
visualized and demonstrated, it nevertheless seems more
appropriate to investigate a system with a single,
well-controlled channel, possibly in a defect- and stress-free
single crystals or in a fully amorphous oxide layer.
Finally, the extinction of the EBIC spots under intense
electron-beam injection and its correlation to a resistance
increase as mentioned above can be understand by the proposed
model. The strong electron injection in a given channel fills all
available traps in the localized band and reduces the intersite
tunneling probability to zero.

\vspace*{-4mm}
\section{Conclusions}

The EBIC technique evidently is of great use for investigating
MIM capacitors with switching and memory properties. First it is
observed that current-leakage sites (white spots) coincide with
local defects in the film. The size of these defects
(nanoditches, impurities, grain boundaries) which is inherent to
film growth, can be very small, typically less than a few hundred
nanometers. Thanks to the enhanced contrast of the EBIC method it
is possible to observe details of the microstructures of the film
below the thin (5-nm) electrode that are not observable by simple
secondary electron SEM imaging. The second observation is that
the switching of the impedance state of the memory cell is
correlated to the number and intensity of the current-leakage
paths. This is demonstrated by a correlated measurement of the
$I$--$V$ characteristics of the device and its EBIC imaging. The
self-similarity of the film's microstructure can explain why
switching and memory effects can in principle be observed in such
heterostructures down to very small length scales. According to
our previous observations, doping with a rather high
concentration of Cr ions (10$^{19}$/cm$^{3}$) seems to help in
stabilizing the switching effect in the perovskites investigated
\cite{Beck}. These ions act as local traps within the insulator,
and in addition to other intrinsic defects form a broad band of
localized states. Switching and nonvolatile memory can well be
explained by a model based on tunneling conduction between
adjacent sites and on the storage or release of charges in traps
at the top of the localized band \cite{Simmons}. The $I$--$V$
dependence proposed in this model is in agreement with our
measurements. There is no reason to doubt that such a mechanism
imagined for the bulk of a material is also valid in the core of
a leakage path in the prebreakdown regime. Therefore it would be
relevant to create such leakage paths with stable and
reproducible switching and memory properties in a controlled
manner if one wishes to manufacture large arrays of crosspoint
memories based on two-terminal devices.

\vspace*{-4mm}
\acknowledgments

The authors wish to thank Ch.\ Gerber for investigating the films
by AFM imaging,  M.\ Tschudy for evaporating the metal
electrodes, and M.\ Despont for introducing us to the
manipulation of the JEOL SEM microscope. Fruitful discussions with
J.\ G.\ Bednorz, R.\ Allenspach and M.\ Roulin as well as the
continuous support of P.\ F.\ Seidler during this work are
gratefully acknowledged.

\end{multicols}

\twocolumn

\newpage

\clearpage

\end{document}